\documentclass[12pt,preprint]{JHEP3}

\usepackage{graphicx}
\def\ie{{\it i.e.}}

\def\lbar{\overline}
\def\nn{\nonumber}

\def\ub{{\bar u}}

\def\db{{\bar d}}
\def\sb{{\bar s}}
\def\bb{{\bar b}}

\newcommand{\erra}[2]{#1\pm #2}
\newcommand{\errb}[3]{#1^{+ #2}_{- #3}}

\title{Flavor symmetry analysis of charmless $B \to VP$ decays}

\author{
Cheng-Wei Chiang\\
Department of Physics and Center for Mathematics and Theoretical Physics,\\
National Central University,
Chungli, Taiwan 320, R.O.C. and\\
Institute of Physics, Academia Sinica, Taipei, Taiwan 115, R.O.C.\\
E-mail: \email{chengwei@phy.ncu.edu.tw}
}
\author{
Yu-Feng Zhou\\
Korea Institute for Advanced Study, Seoul 130-722, Korea \\
E-mail: \email{yfzhou@kias.re.kr}
}

\date{\today}

\preprint{KIAS-P08061}
\abstract{Based upon flavor SU(3) symmetry, we perform global fits to charmless
  $B$ decays into one pseudoscalar meson and one vector meson in the final
  states.  We consider different symmetry breaking schemes and find that the one
  implied by na{\"i}ve factorization is slightly favored over the exact symmetry
  case.  The $(\bar\rho,\bar\eta)$ vertex of the unitarity triangle (UT)
  constrained by our fits is consistent with other methods within errors.  We
  have found large color-suppressed, electroweak penguin and singlet penguin
  amplitudes when the spectator quark ends up in the final-state vector meson.
  Nontrivial relative strong phases are also required to explain the data.  The
  best-fit parameters are used to compute branching ratio and CP asymmetry
  observables in all of the decay modes, particularly those in the $B_s$ decays
  to be measured at the Tevatron and LHC experiments.}

\begin{document}

\section{Introduction \label{sec:intro}}

Thanks to the B-factories, a plethora of data on rare hadronic $B$ meson decays
have become available in recent years.  Because they involve $W$-mediated
charged-currents through mixing and/or decay, these decay modes provide
particular useful information on the CP-violating weak phases and magnitudes of
elements in the Cabibbo-Kobayashi-Maskawa (CKM) matrix
\cite{Cabibbo:1963yz,Kobayashi:1973fv} for the quark sector of the standard
model (SM).  Advances in both experiment and theory have helped us narrow down
these parameters to a high precision.  Through such efforts, it therefore
becomes possible for us to search for evidences of new physics, if any.

Due to the hadronic nature of particles involved in the decays, strong phases
associated with the decay amplitudes that are derived from short-distance
physics as well as final-state interactions are also important.  Even though
they cannot be computed from first principles, these phases play a crucial role
in direct CP asymmetries.  Determination of their pattern and magnitudes in $B$
decays give a test to our knowledge of strong dynamics in the SM.

An approach utilizing the flavor symmetry to relate magnitudes and strong phases
of amplitudes
\cite{Zeppenfeld:1980ex,Savage:1989ub,Chau:1990ay,Gronau:1994rj,Gronau:1995hn}
has been taken to analyze the rare $B$ decay data.  It has the advantage of
reducing model dependence for computing matrix elements of hadronic transitions,
in comparison with the usual perturbative approaches.

In Ref.~\cite{Chiang:2006ih}, we have updated the analysis for $B$ decays into
two charmless pseudoscalar mesons in the final states, and further tested the
flavor symmetry assumption by considering several different breaking schemes in
the amplitudes.  By performing global fits, we find that our results are robust
against fluctuations of individual data with large uncertainties, and different
schemes have roughly the same predictions.

In this article, we concentrate on the rare $B \to VP$ decays, where $V$ and $P$
denote charmless vector and pseudoscalar mesons, respectively.  There have been
some numerical works in the perturbation framework of quantum chromodynamics
(QCD) to calculate the decay rates and CP asymmetries of these decays over the
years.  Na{\"i}ve and general factorization analyses were considered in
Refs.~\cite{Kramer:1994in,Deshpande:1997rr,Ali:1998eb}.  The QCD factorization
(QCDF) method was employed in
Refs.~\cite{Yang:2000xn,Du:2002up,Sun:2002rn,Aleksan:2003qi,Beneke:2003zv,Li:2006jb}.
The calculations using the perturbative QCD approach are scattered in
Refs.~\cite{Lu:2000hj,Chen:2001pr,Liu:2005mm,Guo:2006uq,Guo:2007vw,Ali:2007ff}.
Recently, the soft-collinear effective theory (SCET) was also used in
Ref.~\cite{Wang:2008rk}.  In parallel, some attempts that apply the flavor
symmetry to the $VP$ decays are given in
Refs.~\cite{Dighe:1997wj,Gronau:1999hq,Gronau:2000az,Chiang:2003pm,Wang:2008rk}.

The $B \to VP$ decay modes present a richer structure than the $PP$ final states
because the light spectator quark in $B$ meson can end up in a spin-0 or spin-1
meson, even though the quark-level subprocess is exactly the same.  Moreover,
the number and precision of observables in these modes (particularly the
strangeness-changing ones) have improved considerably in recent years.  Totally,
there are 52 observables in the $VP$ decays.  All the branching ratio and CP
asymmetry observables in the strangeness-changing decays of $B^{0,+}$ mesons
have been measured.  The branching ratio of $\rho^+ K^0$, in particular,
provides valuable information on the magnitude of one type of QCD penguin
amplitude.  In contrast, the observables in the strangeness-conserving
transitions are mostly measured in the $B^+$ decays.  Moreover, some data points
have shifted by noticeable amounts.  For example, the central values of the
branching ratios of $B^+ \to \rho^+ \eta^{(\prime)}$, $B^+ \to K^{*+} \eta$, and
$K^{*+} \pi^-$ have dropped by about $30\%$ from five years ago.  The branching
ratios of $B^0 \to \rho^{\mp} \pi^{\pm}$ also move significantly upward and
downward, respectively.  Therefore, we consider it timely to re-analyze the data
and, at the same time, relax some of the assumptions made in
Ref.~\cite{Chiang:2003pm} in view of the better data pool, and make predictions
for the $B_s$ decay modes which are going to be measured at Tevatron and LHCb.

The structure of this paper is as follows.  In section~\ref{sec:notation}, we
introduce the notation used in our approach and present both measured
observables and amplitude decomposition for the decay modes.  In
section~\ref{sec:fits}, we show our fitting results of the theory parameters in
different schemes.  Discussions and predictions based on our fits are given in
section~\ref{sec:discuss}.  Section~\ref{sec:summary} summarizes our findings in
this work.

\section{Formalism and Notation \label{sec:notation}}

For a two-body $B \to V \, P$ decay process, the magnitude of its invariant
decay amplitude $M$ is related to the partial width in the following way:
\begin{eqnarray}
  \Gamma(B \to V \, P)
  = \frac{|{\bf p}|}{8\pi m_B^2} |M|^2 ~,
\end{eqnarray}
where $\bf p$ is the 3-momentum of the final state particles in the rest frame
of the $B$ meson of mass $m_B$.  To relate partial widths to branching ratios,
we use the world-average lifetimes $\tau^+ = (1.638 \pm 0.011)$ ps, $\tau^0 =
(1.530 \pm 0.009)$ ps, and $\tau_s = (1.437 \pm 0.031)$ ps computed by the Heavy
Flavor Averaging Group (HFAG) \cite{HFAG}.  Each branching ratio quoted in this
paper has been $CP$-averaged.

To perform the flavor amplitude decomposition, we use the following quark
content and phase conventions for mesons:
\begin{itemize}
\item{ {\it Bottom mesons}: $B^0=d\bb$, ${\lbar B}^0=b\db$, $B^+=u\bb$,
    $B^-=-b\ub$, $B_s=s\bb$, ${\lbar B}_s=b\sb$;}
\item{ {\it Pseudoscalar mesons}: $\pi^+=u\db$, $\pi^0=(d\db-u\ub)/\sqrt{2}$,
    $\pi^-=-d\ub$, $K^+=u\sb$, $K^0=d\sb$, ${\lbar K}^0=s\db$, $K^-=-s\ub$,
    $\eta=(s\sb-u\ub-d\db)/\sqrt{3}$,
    $\eta^{\prime}=(u\ub+d\db+2s\sb)/\sqrt{6}$;}
\item{ {\it Vector mesons}: $\rho^+=u\db$, $\rho^0=(d\db-u\ub)/\sqrt{2}$,
    $\rho^-=-d\ub$, $\omega=(u\ub+d\db)/\sqrt{2}$, $K^{*+}=u\sb$,
    $K^{*0}=d\sb$, ${\lbar K}^{*0}=s\db$, $K^{*-}=-s\ub$, $\phi=s\sb$.}
\end{itemize}
The $\eta$ and $\eta'$ mesons correspond to octet-singlet mixtures
\begin{eqnarray}
\eta  &=& \eta_8 \cos \theta_0 - \eta_1 \sin \theta_0 ~, \\
\eta' &=& \eta_8 \sin \theta_0 + \eta_1 \cos \theta_0 ~.
\end{eqnarray}
As shown in Ref.~\cite{Chiang:2003pm}, varying the mixing angle $\theta_0$ does
not improve the quality of fits.  For convenience, we fix $\theta_0 =
\sin^{-1}(1/3) \simeq 19.5^\circ$ according to the above-mentioned quark
contents of $\eta$ and $\eta'$.

We list flavor amplitude decompositions and averaged experimental data for $B
\to VP$ decays in Tables \ref{tab:VPdS0} and \ref{tab:VPdS1}.  Values of
measured observables are obtained from the 
latest 2008 summer results of the HFAG
\cite{HFAG}.

\TABLE{\small
  \caption{Flavor amplitude decomposition and measured
    observables~\cite{BABARExp,BelleExp,CDFExp,CLEOExp} of
    strangeness-conserving $B \to VP$ decays.  The time-dependent CP asymmetries
    $\cal A$ and $\cal S$, if applicable, are listed in the first and second
    rows, respectively.}
\label{tab:VPdS0}
\begin{tabular}{|llccc|}
\hline
\multicolumn{2}{|c}{Mode} & Flavor Amplitude
 & BR ($\times 10^{-6}$) & $A_{CP}$ \\ 
\hline
$B^+ \to$
    & ${\lbar K}^{*0} K^+$
        & $p_P$
        & $0.68 \pm 0.19$ & - \\
    & $K^{*+} \lbar K^0$
        & $p_V$
        & - & - \\
    & $\rho^0 \pi^+$ 
        & $-\frac{1}{\sqrt{2}}(t_V+c_P+p_V-p_P)$
        & $\errb{8.7}{1.0}{1.1}$ & $\errb{-0.07}{0.12}{0.13}$ \\
    & $\rho^+ \pi^0$ & $-\frac{1}{\sqrt{2}}(t_P+c_V+p_P-p_V)$
        & $\errb{10.9}{1.4}{1.5}$ & $\erra{0.02}{0.11}$ \\
    & $\rho^+ \eta$ 
        & $-\frac{1}{\sqrt{3}}(t_P+c_V+p_P+p_V+s_V)$
        & $\erra{6.9}{1.0}$ & $\erra{0.11}{0.11}$ \\
    & $\rho^+ \eta'$ 
        & $\frac{1}{\sqrt{6}}(t_P+c_V+p_P+p_V+4s_V)$
        & $\errb{9.1}{3.7}{2.8}$ & $\erra{-0.04}{0.28}$ \\
    & $\omega \pi^+$ 
        & $\frac{1}{\sqrt{2}}(t_V+c_P+p_P+p_V+2s_P)$
        & $\erra{6.9}{0.5}$ & $\erra{-0.04}{0.06}$ \\
    & $\phi \pi^+$
        & $s_P$
        & $<0.24$ & - \\
\hline
$B^0 \to$
    & $\overline{K}^{*0} K^0$
        & $p_P$ 
        & - & - \\
    & $K^{*0} \overline{K}^0$
        & $p_V$
        & $<1.9$ & - \\
    & $\rho^- \pi^+$
        & $-(t_V+p_V)$
        & $16.42 \pm 1.96^a$ & $0.12 \pm 0.06^a$ \\
    & & & & $-0.04 \pm 0.13^a$ \\
    & $\rho^+ \pi^-$
        & $-(t_P+p_P)$
        & $7.58 \pm 1.25^a$ & $-0.14 \pm 0.12^a$ \\
    & & & & $0.06 \pm 0.13^a$ \\
    & $\rho^0 \pi^0$ 
        & $-\frac12(c_P+c_V-p_P-p_V)$
        & $\erra{2.0}{0.5}$ & - \\
    & $\rho^0 \eta$ 
        & $\frac{1}{\sqrt{6}}(c_P-c_V-p_P-p_V-s_V)$
        & $<1.5$ & - \\
    & $\rho^0 \eta'$ 
        & $-\frac{1}{2\sqrt{3}}(c_P-c_V-p_P-p_V-4s_V)$
        & $<1.3$ & - \\
    & $\omega \pi^0$
        & $\frac{1}{2}(c_P-c_V+p_P+p_V+2s_P)$
        & $<0.5$ & - \\
    & $\omega \eta$
        & $-\frac{1}{\sqrt{6}}(c_P+c_V+p_P+p_V+2s_P+s_V)$
        & $< 1.6$ & - \\
    & $\omega \eta'$
        & $\frac{1}{2\sqrt{3}}(c_P+c_V+p_P+p_V+2s_P+4s_V)$
        & $< 1.9$ & - \\
    & $\phi \pi^0$
        & $\frac{1}{\sqrt{2}}s_P$
        & $<0.28$ & - \\
    & $\phi \eta$
        & $-\frac{1}{\sqrt{3}}s_P$
        & $< 0.52$ & - \\
    & $\phi \eta'$
        & $\frac{1}{\sqrt{6}}s_P$
        & $ < 0.5$ & - \\
\hline
$B_s \to$
    & $\overline{K}^{*0} \pi^0$
        & $-\frac{1}{\sqrt{2}}(c_V - p_V)$ 
        & - & - \\
    & $K^{*-} \pi^+$
        & $-(t_V + p_V)$ 
        & - & - \\
    & $\rho^- K^+$
        & $-(t_P + p_P)$ 
        & - & - \\
    & $\rho^0 \overline{K}^0$
        & $-\frac{1}{\sqrt{2}}(c_P - p_P)$ 
        & - & - \\
    & $\overline{K}^{*0} \eta$
        & $-\frac{1}{\sqrt{3}}(c_V - p_P + p_V + s_V)$ 
        & - & - \\
    & $\overline{K}^{*0} \eta'$
        & $\frac{1}{\sqrt{6}}(c_V + 2p_P + p_V + 4 s_V)$ 
        & - & - \\
    & $\omega \overline{K}^0$
        & $\frac{1}{\sqrt{2}}(c_P + p_P + 2s_P)$ 
        & - & - \\
    & $\phi \overline{K}^0$
        & $p_V + s_P$ 
        & - & - \\
\hline
\end{tabular}
$^a$ Values obtained using the method described in Ref.~\cite{Chiang:2003pm}.
}

\TABLE{\small
  \caption{Flavor amplitude decomposition and measured
    observables~\cite{BABARExp,BelleExp,CDFExp,CLEOExp} of strangeness-changing
    $B \to VP$ decays.  The time-dependent CP asymmetries $\cal A$ and $\cal S$,
    if applicable, are listed in the first and second rows, respectively.}
\label{tab:VPdS1}
\begin{tabular}{|llccc|}
\hline
\multicolumn{2}{|c}{Mode} & Flavor Amplitude
 & BR ($\times 10^{-6}$) & $A_{CP}$ \\ 
\hline
$B^+ \to$
    & $K^{*0} \pi^+$
        & $p'_P$
        & $\erra{10.0}{0.8}$ & $\errb{-0.020}{0.057}{0.061}$ \\
    & $K^{*+} \pi^0$ 
        & $-\frac{1}{\sqrt{2}}(t'_P+c'_V+p'_P)$
        & $\erra{6.9}{2.3}$ & $\erra{0.04}{0.29}$ \\
    & $\rho^0 K^+$
        & $-\frac{1}{\sqrt{2}}(t'_V+c'_P+p'_V)$
        & $\errb{3.81}{0.48}{0.46}$ & $\errb{0.417}{0.081}{0.104}$ \\
    & $\rho^+ K^0$
        & $p'_V$
        & $\errb{8.0}{1.5}{1.4}$ & $\erra{-0.12}{0.17}$ \\
    & $K^{*+} \eta$ 
        & $-\frac{1}{\sqrt{3}}(t'_P+c'_V+p'_P-p'_V+s'_V)$
        & $\erra{19.3}{1.6}$ & $\erra{0.02}{0.06}$ \\
    & $K^{*+} \eta'$ 
        & $\frac{1}{\sqrt{6}}(t'_P+c'_V+p'_P+2p'_V+4s'_V)$
        & $\errb{4.9}{2.1}{1.9}$ & $\errb{0.30}{0.33}{0.37}$ \\
    & $\omega K^+$ 
        & $\frac{1}{\sqrt{2}}(t'_V+c'_P+p'_V+2s'_P)$ 
        & $\erra{6.7}{0.5}$ & $\erra{0.02}{0.05}$ \\
    & $\phi K^+$
        & $p'_P+s'_P$ 
        & $\erra{8.30}{0.65}$ & $\erra{0.034}{0.044}$ \\
\hline
$B^0 \to$
    & $K^{*+} \pi^-$
        & $-(t'_P+p'_P)$ 
        & $\erra{10.3}{1.1}$ & $\erra{-0.25}{0.11}$ \\
    & $K^{*0} \pi^0$
        & $\frac{1}{\sqrt{2}}(c'_V - p'_P)$
        & $2.4\pm0.7$ & $-0.15 \pm 0.12$ \\
    & $\rho^- K^+$
        & $-(t'_V+p'_V)$
        & $\errb{8.6}{0.9}{1.1}$ & $\erra{0.15}{0.06}$ \\
    & $\rho^0 K^0$
        & $-\frac{1}{\sqrt{2}}(c'_P - p'_V)$
        & $\errb{5.4}{0.9}{1.0}$ & $\erra{-0.02}{0.29}$ \\
                                &&&& $\erra{0.61}{0.26}$\\
    & $K^{*0} \eta$
        & $-\frac{1}{\sqrt{3}}(c'_V + p'_P-p'_V + s'_V)$
        & $\erra{15.9}{1.0}$ & $\erra{0.19}{0.05}$ \\
    & $K^{*0} \eta'$
        & $\frac{1}{\sqrt{6}}(c'_V + p'_P + 2p'_V + 4s'_V)$
        & $\erra{3.8}{1.2}$ & $\erra{-0.08}{0.25}$ \\
    & $\omega K^0$
        & $\frac{1}{\sqrt{2}}(c'_P + p'_V + 2s'_P)$
        & $\erra{5.0}{0.6}$ & $\erra{0.32}{0.17}$ \\
	                  &&&& $\erra{0.45}{0.24}$\\
    & $\phi K^0$
        & $p'_P+s'_P$ 
        & $\errb{8.3}{1.2}{1.0}$ & $\erra{0.23}{0.15}$ \\
	                       &&&& $\errb{0.44}{0.17}{0.18}$\\
\hline
$B_s \to$
    & $K^{*+} K^-$
        & $-(p'_P + t'_P)$
        & - & - \\
    & $K^{*-} K^+$
        & $-(p'_V + t'_V)$
        & - & - \\
    & $K^{*0} \overline{K}^0$
        & $p'_P$
        & - & - \\
    & $\overline{K}^{*0} K^0$
        & $p'_V$
        & - & - \\
    & $\rho^0 \eta$
        & $-\frac{1}{\sqrt{6}}c'_P$
        & - & - \\
    & $\rho^0 \eta'$
        & $-\frac{1}{\sqrt{3}}c'_P$
        & - & - \\
    & $\omega \eta$
        & $-\frac{1}{\sqrt{6}}(c'_P + 2s'_P)$
        & - & - \\
    & $\omega \eta'$
        & $-\frac{1}{\sqrt{3}}(c'_P + 2s'_P)$
        & - & - \\
    & $\phi \pi^0$
        & $-\frac{1}{\sqrt{2}}c'_V$
        & - & - \\
    & $\phi \eta$
        & $\frac{1}{\sqrt{3}}(p'_P + p'_V - c'_V + s'_P - s'_V)$
        & - & - \\
    & $\phi \eta$
        & $\frac{1}{\sqrt{6}}(2p'_P + 2p'_V + c'_V + 2s'_P + 4s'_V)$
        & - & - \\
\hline
\end{tabular}
}

In the present approximation, we consider only five dominant types of
independent amplitudes: a ``tree'' contribution $T$; a ``color-suppressed''
contribution $C$; a ``QCD penguin'' contribution $P$; a ``flavor-singlet''
contribution $S$, and an ``electroweak (EW) penguin'' contribution $P_{EW}$.
The first four types are considered as the leading-order amplitudes, while the
last one is higher-order in weak interactions.  Depending upon which final state
meson the spectator quark in the $B$ meson ends up in, we further associate a
subscript $P$ or $V$ to the above-mentioned amplitudes.  For example, $T_P$ and
$T_V$ denote a tree amplitude with the spectator quark of the $B$ meson going
into the pseudoscalar and vector meson in the final state, respectively.  These
two kinds of amplitudes are different in general.  In the following, we will
suppress the subscripts $P,V$ when discussions apply to both classes of
amplitudes of each type.

There are also other types of amplitudes, such as the ``color-suppressed EW
penguin'' diagram $P_{EW}^C$, ``exchange'' diagram $E$, ``annihilation'' diagram
$A$, and ``penguin annihilation'' diagram $PA$.  Due to dynamical suppression,
these amplitudes are ignored in the analysis.

The QCD penguin amplitude contains three components (apart from the CKM
factors): $P_t$, $P_c$, and $P_u$, with the subscript denoting which quark is
running in the loop.  After imposing the unitarity condition, we can remove the
explicit $t$-quark dependence and are left with two components: $P_{tc} = P_t -
P_c$ and $P_{tu} = P_t - P_u$.  For simplicity, we assume the $t$-penguin
dominance, so that $P_{tc} = P_{tu} \equiv P$.  The same comment applies to the
EW penguin and singlet penguin amplitudes, too.

In physical processes, the above-mentioned flavor amplitudes always appear in
specific combinations.  To simplify the notations, we therefore define the
following unprimed and primed symbols for $\Delta S = 0$ and $|\Delta S| = 1$
transitions, respectively:
\begin{eqnarray}
\label{eqn:dict}
t \equiv Y_{db}^u T - (Y_{db}^u + Y_{db}^c) P_{EW}^C ~,
&\quad&
t' \equiv Y_{sb}^u \xi_t T - (Y_{sb}^u + Y_{sb}^c) P_{EW}^C ~, \nn \\
c \equiv Y_{db}^u C - (Y_{db}^u + Y_{db}^c) P_{EW} ~,
&\quad&
c' \equiv Y_{sb}^u \xi_c C - (Y_{sb}^u + Y_{sb}^c) P_{EW} ~, \nn \\
p \equiv - (Y_{db}^u + Y_{db}^c) \left( P - \frac{1}{3} P_{EW}^C \right) ~,
&\quad&
p' \equiv - (Y_{sb}^u + Y_{sb}^c) \left( \xi_p P - \frac{1}{3} P_{EW}^C \right)
~, \nn
\\
s \equiv - (Y_{db}^u + Y_{db}^c) \left( S - \frac{1}{3} P_{EW} \right) ~,
&\quad&
s' \equiv - (Y_{sb}^u + Y_{sb}^c) \left( \xi_s S - \frac{1}{3} P_{EW} \right)
~,
\end{eqnarray}
where $Y_{qb}^{q'} \equiv V_{q'q} V_{q'b}^*$ ($q \in \{ d,s \}$ and $q' \in \{
u,c \}$).  Here we also keep the $P_{EW}^C$ amplitude for completeness, though
it is ignored in the subsequent analysis.  Again, all the above amplitudes are
to be associated with subscript $P$ or $V$, depending on the process.  Here we
have explicitly factored out the CKM factors, but leave strong phases inside the
amplitudes.

From $\Delta S = 0$ to $|\Delta S| = 1$ transitions, we put in SU(3) breaking
factors $\xi_{T_{P,V}}, \xi_{C_{P,V}}$, and $\xi_{P_{P,V}}$ for $T_{P,V}$,
$C_{P,V}$, and $P_{P,V}$, respectively.  If some type of amplitudes is
factorizable, the corresponding SU(3) breaking factor is either $f_K/f_{\pi} =
1.22$ or $f_{K^*} / f_{\rho} = 1.00$ \cite{PDG}.  For example, we have for the
$B^0 \to K^{*+}\pi^-$ decay:
\begin{eqnarray*}
  \mathcal{A}(K^{*+}\pi^-) &=&
  - Y_{sb}^{u} \xi_t T_P + \left( Y_{sb}^{u} + Y_{sb}^{c} \right) \xi_p P_P ~.
\end{eqnarray*}
This can be obtained from the complete set of flavor amplitude decomposition
given in Table \ref{tab:VPdS1}, Table \ref{tab:VPdS1} and appropriate forms of
Eqs.~(\ref{eqn:dict}).

In this analysis, the CKM factors are expressed in terms of the Wolfenstein
parameterization \cite{Wolfenstein:1983yz} to $O(\lambda^5)$.  Since $\lambda$
has been determined from kaon decays to a high accuracy, we will use the
central value $0.2272$ quoted by the CKMfitter group \cite{CKMfitter} as a
theory input, and leave $A$, $\bar\rho \equiv \rho ( 1 - \lambda^2/2 )$, and
$\bar\eta \equiv \eta ( 1 - \lambda^2/2 )$ as fitting parameters to be
determined by data.

For the $B$ meson decaying into a CP eigenstate $f_{CP}$, the time-dependent CP
asymmetry is written as
\begin{eqnarray}
\label{eq:t-dep}
A_{CP}(t)
&=& 
\frac{\Gamma(\bar{B}^{0}\to f_{CP})-\Gamma(B^{0}\to f_{CP})}
     {\Gamma(\bar{B}^{0}\to f_{CP})+\Gamma(B^{0}\to f_{CP})} \nn \\
&=&
{\cal S} \, \sin(\Delta m_{B} \cdot t) 
+ {\cal A} \, \cos(\Delta m_{B} \cdot t) ~,
\end{eqnarray}
where $\Delta m_{B}$ is the mass difference between the two mass eigenstates of
$B$ mesons and $t$ is the decay time measured from the tagged $B$ meson.

\section{Fitting Analysis \label{sec:fits}}

In this section, we present the following two schemes in our fits:
\begin{enumerate}
\item exact flavor symmetry for all amplitudes (\ie, $\xi_{T_{P,V}} = \xi_{C_{P,V}}
  = \xi_{P_{P,V}} = 1$);
\item imposing partial SU(3)-breaking factors on $T$ and $C$ amplitudes only
  (\ie, $\xi_{T_P,C_P} = f_{K^*} / f_{\rho}$ and $\xi_{T_V,C_V} = f_K / f_\pi$,
  while $\xi_{P_P,P_V} = 1$);
\end{enumerate}
We have assumed exact flavor symmetry for the strong phases to reduce
independent parameters in our fits.  Besides, $T_P$ is fixed to be real and
positive in our phase convention (\ie, $\delta_{T_P} = 0$).  All the other
strong phases are measured with respect to it.

We further divide our fits into two classes: (A) the $VP$ modes that do not
involve singlet penguin contributions, and (B) all of the $VP$ modes.  As shown
in Table \ref{tab:VPdS0} and Table \ref{tab:VPdS1}, the modes that contain the
singlet penguin amplitudes are those having $\eta$, $\eta'$, $\phi$, or $\omega$
in the final states.

It is appropriate to list some major differences between the current analysis
and Ref.\cite{Chiang:2003pm}.  Throughout this analysis, we do not assume any
strong phase relation between the EW penguin, singlet penguin, and the QCD
penguin amplitudes.  Neither do we assume any strong phase relation between the
color-suppressed amplitudes and the tree amplitudes.  The relative size and
phase of $P_P$ and $P_V$ are always kept free.  Moreover, we do not assume $S_P$
to be small enough for omission.  Instead, we keep and constrain its magnitude
and phase.

In the following, we perform $\chi^2$ fits to the observables in the $B \to VP$
modes as well as $|V_{ub}| = (4.26 \pm 0.36) \times 10^{-3}$ and $|V_{cb}| =
(41.63 \pm 0.65) \times 10^{-3}$ \cite{CKMfitter} for the above-mentioned two
schemes.  The inclusion of $|V_{ub}|$ and $|V_{cb}|$ helps fixing the values of
$A$ and $\sqrt{{\bar\rho}^2 + {\bar\eta}^2}$.  However, we drop the branching
ratio and direct CP asymmetry of the $B^0 \to K^{*0} \pi^0$ decay from the fits
because currently the BABAR Collaboration and the Belle Collaboration have a
large disagreement in the branching ratio, whose weighted average is $(2.42 \pm
1.16) \times 10^{-6}$ with a scale factor $S = 1.77$.  As we will see later, our
predictions based on best fits deviate much from these two observables.

The fit results of theory parameters are summarized in Table~\ref{tab:Fit}.  As
given in the table, Scheme 1 of exact SU(3) symmetry is slightly worse than
Scheme 2.  As defined above, the main difference between these two schemes is in
the scaling behavior of $T_P$ and $C_P$ between the strangeness-conserving and
strangeness-changing modes.  We have also tried other schemes, such as having
additional symmetry breaking for amplitude sizes.  However, either the fitting
quality becomes worse or they involve too large SU(3) breaking (over 30\%).  We
will present our plots and predictions mainly for Scheme 2.

\TABLE{
  \caption{Fit results (1-$\sigma$ ranges) of the theory parameters for Classes
    (A) and (B) in the two schemes defined in the text.  The minimal $\chi^2$
    value and the number of degrees of freedom (dof) are also given.  The
    amplitudes are given in units of $10^4$ eV, and the phases are in degrees.}
\label{tab:Fit}
\begin{tabular}{|lllll|}
\hline
Parameter & 
\multicolumn{4}{c|}{Scheme} \\
& 1A & 2A & 1B & 2B \\
\hline
$T_P         $               & $0.721\pm0.088$                       & $0.727\pm0.089$                                & $0.785\pm0.098$                                   & $0.791\pm0.100$                                
\\                 	                                                             $T_V         $     	     & $1.069^{+0.119}_{-0.104}$             & $1.070^{+0.119}_{-0.105}$                      & $1.168^{+0.131}_{-0.116}$                         & $1.170^{+0.133}_{-0.118}$                       
\\                 	                                                             $\delta_{T_V}$     	     & $1.9\pm5.7$                       & $2.3\pm5.7$                                & $0.4\pm5.5$                                   & $0.6\pm5.4$                                 
\\                 	                                                             $C_P         $     	     & $0.093^{+0.209}_{-0.253}$             & $0.184\pm0.223$                                & $0.173^{+0.138}_{-0.108}$                         & $0.122^{+0.125}_{-0.089}$                       
\\                 	                                                             $\delta_{C_P}$     	     & $-118.9\pm77.4$                   & $-107.7\pm31.0$                            & $133.0\pm34.2$                                & $149.0^{+72.4}_{-43.0}$                   
\\                 	                                                             $C_V         $     	     & $0.688^{+0.226}_{-0.174}$             & $0.624^{+0.209}_{-0.154}$                      & $0.945\pm0.142$                                   & $0.892\pm0.139$                                 
\\                 	                                                             $\delta_{C_V}$     	     & $-66.0^{+30.3}_{-22.7}$         & $-57.0^{+31.3}_{-25.2}$                  & $-82.0^{+12.0}_{-10.1}$                     & $-75.9^{+12.6}_{-10.7}$                   
\\                 	                                                             $P_P         $         	     & $0.084\pm0.003$                       & $0.084\pm0.003$                                & $0.085\pm0.003$                                   & $0.085\pm0.003$                                 
\\                    	                                                             $\delta_{P_P}$     	     & $-3.9\pm10.2$                     & $-5.7\pm10.0$                               & $-1.0\pm8.0$                                  & $-2.6\pm7.8$                                
\\                 	                                                             $P_V         $     	     & $0.065\pm0.004$                       & $0.063\pm0.004$                                & $0.068\pm0.004$                                   & $0.066\pm0.004$                                 
\\                 	                                                             $\delta_{P_V}$     	     & $171.7\pm8.1$                     & $172.6\pm7.7$                              & $172.2\pm7.1$                                 & $172.5\pm6.9$                               
\\                 	                                                             $P_{EW,P}    $     	     & $0.039^{+0.009}_{-0.011}$             & $0.039^{+0.009}_{-0.010}$                      & $0.032^{+0.010}_{-0.013}$                         & $0.031^{+0.010}_{-0.011}$                       
\\                 	                                                             $\delta_{P_{EW,P}}$	     & $56.4^{+10.4}_{-11.6}$          & $55.1^{+10.4}_{-11.9}$                   & $60.9^{+10.0}_{-15.1}$                      & $59.0^{+10.5}_{-15.8}$                    
\\                 	                                                             $P_{EW,V}    $     	     & $0.067\pm0.049$                       & $0.052^{+0.048}_{-0.041}$                      & $0.096^{+0.027}_{-0.030}$                         & $0.087\pm0.029$                                 
\\
$\delta_{P_{EW,V}}$	     & $-98.7^{+52.0}_{-23.3}$         &
$-90.2^{+82.0}_{-26.1}$                  & $-113.5^{+9.6}_{-8.2}$
& $-111.0^{+10.4}_{-8.6}$     \\
$S_P         $     	     & fixed                                 & fixed                                          & $0.015^{+0.005}_{-0.005}$                         & $0.014^{+0.004}_{-0.004}$                       
\\                 	                                                             $\delta_{S_P}$     	     & fixed                                 & fixed                                          & $-133.4^{+16.0}_{-23.9}$                    & $-139.8^{+16.6}_{-23.5}$                   
\\                 	                                                             $S_V         $     	     & fixed                                 & fixed                                          & $0.049\pm0.005$                                   & $0.048\pm0.005$                                 
\\                 	                                                             $\delta_{S_V}$     	     & fixed                                 & fixed                                          & $-49.4^{+22.2}_{-18.6}$                     & $-47.7^{+21.5}_{-18.3}$                   
\\                 	                                                             \hline
$A            $    	     & $0.807\pm0.013$                       & $0.807\pm0.013$                                & $0.809\pm0.012$                                   & $0.809\pm0.012$                                 
\\
$\bar{\rho}   $    	     & $0.151\pm0.036$                       & $0.146\pm0.035$                                & $0.116\pm0.030$                                   & $0.109^{+0.030}_{-0.028}$                       
\\
$\bar{\eta}   $    	     & $0.401\pm0.030$                       & $0.400\pm0.030$                                & $0.373\pm0.029$                                   & $0.371\pm0.030$                                 
\\
\hline 
$\chi^{2}/dof     $    	     & $20.7/8$                                & $19.9/8$                                         & $44.6/30$                                            & $44.5/30$                                          
\\
\hline
\end{tabular}
}

Some general features are observed in these fits.  The two types of tree
amplitude have roughly the same strong phases, with $T_V$ larger than $T_P$ by
about $50\%$, largely driven by the branching ratios of $\rho^{\mp} \pi^{\pm}$.
The $C_V$ amplitude is 3 to 7 times larger than the $C_P$ amplitude.  Both of
them have sizeable strong phases relative to the tree amplitudes.  Moreover, the
strong phases of $C_P$ changes abruptly when we enlarge our fitting set from
Class (A) to Class (B).  They are correlated because they appear in combination
in the physical amplitudes.

The best fitted ratios between color-suppressed tree and tree amplitudes are
\begin{eqnarray}\label{C:T}
\begin{tabular}{c c c c c}
    & (1A) & (2A) & (1B) & (2B) \\
$C_V/T_V=$ 	&  $ 0.64 \pm 0.20$ 	& $0.58 \pm 0.18$  	& $0.81 \pm
0.15$ & $0.76 \pm 0.14$\\
$C_P/T_P=$	&  $0.13 \pm 0.32$	& $0.25 \pm 0.31$	& $0.22 \pm
0.16$ & $0.15 \pm 0.16$ .\\ 
\end{tabular}
\end{eqnarray}
In the four schemes, the central values of the ratio $C_P/T_P $ range from
$0.13$ to $0.25$, agreeing with our na{\"i}ve expectation, even though one still
cannot take them seriously due to the large errors coming from the uncertainty
in $C_P$.  On the other hand, the central values for $C_V/T_V$ are significantly
larger with less uncertainties.  The value of $C_V$ increases by about $40\%$
from Set (A) to Set (B) though.  The four schemes favor $C_V/T_V$ in the range
of $0.58\sim 0.76$.  As a comparison, the default parameter set of the QCDF
approach \cite{Beneke:2003zv} gives
\begin{eqnarray}
C_V/T_V & = &0.158 \pm 0.109\mbox{ and } C_P/T_P =  0.20 \pm 0.13 ,  
\end{eqnarray}
The large $C_V/T_V$ ratio is close to what we have found for $C/T \sim 0.65$ in
the $B$ decays to two pseudoscalars
\cite{Chiang:2004nm,Barshay:2004hb,Wu:2004xx,Charng:2004ed,He:2004ck,Wu:2005hi,%
  Chiang:2006ih,Wu:2007tm}.  Even though such values of $C/T$ in the $PP$ decays
and $C_V/T_V$ pose a challenge to perturbative calculations, they seem to follow
the simple pattern of factorization in tree and color-suppressed tree
amplitudes.

The best fitted ratios between the QCD penguin amplitudes and the tree
amplitudes are pretty stable among different schemes considered in this
work.  The ratios for the four schemes are given by
\begin{eqnarray}\label{P:T}
\begin{tabular}{c c c c c}
    & (1A) & (2A) & (1B) &(2B) \\
$P_V/T_V=$              &  $ 0.06 \pm 0.01$  &  $ 0.06 \pm 0.01$  &  $ 0.06 \pm
0.01$&  $ 0.06 \pm 0.01$\\
$P_P/T_P =$               & $ 0.12 \pm 0.01$  &  $ 0.12 \pm 0.01$  & $0.11 \pm
0.01$& $0.11 \pm 0.01$ . \\
\end{tabular}
\end{eqnarray}
In comparison, the ratio $P/T \sim 0.21$ in the $PP$ modes \cite{Chiang:2006ih}.
The strong phase of $P_P$ is the same as $T_P$ within a few degrees, whereas
that of $P_V$ is about $180^\circ$ different.  This agrees the expectation of
Refs.~\cite{Lipkin:1997ad,Lipkin:1998ew,Lipkin:2007yi} and reassures our
previous finding \cite{Chiang:2003pm} using old data.  However, it is worth
noting that in this work this solution is found even without invoking the $B \to
K^* \eta$ decays.  The QCDF default values are \cite{Beneke:2003zv}
\begin{eqnarray}
P_V/T_V & = &0.035\pm 0.017 \mbox{ and } P_P/T_P = 0.032\pm 0.006 ~,
\end{eqnarray}
and also favors an opposite phase between $P_P$ and $P_V$ amplitudes.  Such a
phase difference is due to the chiral enhancement that results in a sign flip in
the effective coefficients for the QCD penguin amplitudes.  Note, however, that
the magnitudes of the QCD penguin amplitudes derived in QCDF are significantly
smaller than what we find.  It has been noticed that they cannot account for
some large branching ratios in the QCD penguin-dominated modes
\cite{Aleksan:2003qi}.

The strong phase between $P_P$ and $P_V$ is about $180^\circ$, with the former
roughly in phase with $T_P$.  Such a phase difference produces maximal
constructive or destructive interference effects in decay modes that involve
both of them.  Since the relative phases among the tree- and penguin-type
amplitudes are trivial (\ie, $\sim 0^\circ$ or $180^\circ$), as will be seen
later, we generally do not expect large direct CP asymmetries in the decay modes
involving only them.

For the EW penguin amplitudes, the constraint on $P_{EW,P}$ is better than
$P_{EW,V}$ in Set (A).  Nevertheless, the constraint on $P_{EW,V}$ improves in
Set (B).  We note that the strong phases of $P_{EW,P}$ and $P_{EW,V}$ are
significantly different from those of $P_P$ and $P_V$, unlike the assumption
made in Ref.~\cite{Chiang:2003pm}.  It is interesting to notice that $P_{EW,V}$
increases by about $50\%$ from fits of Set (A) to fits of Set (B).  At the same
time, the uncertainty in the strong phase associated with $P_{EW,V}$ improves.
In Set (B), $P_{EW,V}$ is about 3 times larger than $P_{EW,P}$.  To one's
surprise, $P_{EW,V}$ is unexpectedly large, in line with $C_V$.

As to the singlet penguin amplitudes, we find that $S_P$ is about 3 times
smaller than $S_V$.  This partly justifies the ignorance of the former made in
Ref.~\cite{Chiang:2003pm}, in view of the Okubo-Zweig-Iizuka (OZI) rule.
Moreover, if one compares the central values, the $S_P$ amplitude has a strong
phase in roughly the opposite direction of $P_P$ and subtends a nontrivial angle
from $C_P$.  The $S_V$ amplitude has a $\sim 220^\circ$ phase shift from $P_V$
and deviates from $C_V$ by about $30^\circ$.  It is interesting to note that the
physical amplitude $s_P$ has a completely constructive interference between
$S_P$ and $P_{EW,P}/3$.  Also, both types of singlet penguin amplitudes are
about half the sizes of the corresponding EW penguin amplitudes.

Here we describe qualitatively how some of the theory parameters are fixed by
data, thereby explaining their associated uncertainties.  For this, we
temporarily concentrate on the modes without involving singlet penguin
amplitudes.  But the argument can be easily extended to all modes.  In our fits,
the determination of $P_P$ and $P_V$ is most precise because they can be
directly extracted from the strangeness-changing $B^+ \to K^{*0} \pi^+$ and
$\rho^+ K^0$ modes.  The next precise parameters are the magnitudes of tree
amplitudes and their phase shifts relative to the QCD penguin amplitudes.  They
are fixed mainly by the strangeness-conserving $B^0 \to \rho^\pm \pi^\mp$ and to
some extent by the strangeness-changing $B^0 \to \rho^- K^+$ and $K^{*+} \rho^-$
modes.  Since no direct CP asymmetry is observed in these modes, the relative
strong phases are seen to be trivial.

As the color-suppressed and EW penguin amplitudes of the same type (subscript
$P$ or $V$) always show up in pairs in the physical processes, the determination
of their sizes and strong phases becomes trickier.  This is because the
color-suppressed amplitudes dominate in the $\Delta S = 0$ processes, whereas
the EW penguin amplitudes play more role in the $|\Delta S| = 1$ decays.  This
explains why $C_V$ is better determined whereas $P_{EW,V}$ is not, for ${\cal
  B}(B^+ \to \rho^+ \pi^0)$ is more precise than ${\cal B}(K^{*+} \pi^0)$.
Likewise, the precision on $C_P$ is worse than $P_{EW,P}$ because the
combination of ${\cal B}(B^+ \to \rho^0 K^+)$ and ${\cal B}(B^0 \to \rho^0 K^0)$
is better than ${\cal B}(\rho^0 \pi^+)$.

Since the singlet penguin amplitudes are loop-mediated, they are better
constrained by the $|\Delta S| = 1$ decay modes.  Currently, both charged and
neutral $\phi K$ modes have consistent branching ratios and direct CP
asymmetries.  This basically fixes the magnitude and phase of $S_P$.  In
contrast, $S_V$ is constrained in a more involved manner through interference
with other amplitudes.

We note in passing that in Class (A), we have also found other sets of
parameters that render smaller $\chi^2_{\rm min}$ in the fits.  They are not
listed in the tables because they are not favored once the modes involving the
singlet penguin amplitudes are taken into account.  A distinctive feature of
such solutions from the above-mentioned ones is that either the relative strong
phase between $P_P$ and $P_V$ is close to zero or that between $T_P$ and $T_V$
is close to $180^\circ$.  In the former case, an interesting feature is that the
ratios $C_P/T_P = 0.57 \pm 0.43$ and $C_V/T_V = 0.49 \pm 0.12$ in Scheme 2.
They become comparable to each other, but still much larger than the usual
perturbative expectation.  In the latter case, we obtain a somewhat small $\rho
= 0.08$.

In Fig.~\ref{fig:rhoeta-scheme1}, we show the contours of the $(\bar\rho,
\bar\eta)$ vertex at the 1-$\sigma$ and 95\% confidence level (CL) obtained
using Scheme 2.  The left plot uses a fit to modes without involving the singlet
penguin amplitudes.  In this case, our favored region of the vertex is slightly
higher than that given by the CKMfitter \cite{CKMfitter} and UTfit \cite{UTfit}.
The right plot is a global fit to all the $VP$ modes.  Comparing to the left
plot, we see that the favored region shifts lower and to the left on the
$\bar\rho$-$\bar\eta$ plane.  In this case, the preferred value of $\beta$
agrees with other methods, while the value of $\gamma$ is slightly larger.  The
best fitted three angles in the UT are
\begin{eqnarray}
\alpha=(83\pm8)^\circ  & \mbox{ or } & 72^\circ<\alpha<99^\circ (95\% {\rm CL}) ~,
\nonumber\\
\beta=(26\pm2)^\circ   & \mbox{ or } & 18^\circ<\beta<30^\circ (95\% {\rm CL}) ~,
\nonumber\\
\gamma=(71\pm5)^\circ  & \mbox{ or } & 62^\circ<\gamma<78^\circ (95\% {\rm CL})
\end{eqnarray}
for Scheme (2A), and
\begin{eqnarray}
\alpha=(84\pm6)^\circ  & \mbox{ or } & 77^\circ<\alpha<95^\circ (95\% {\rm CL}) ~,
\nonumber\\
\beta=(23\pm2)^\circ   & \mbox{ or } & 18^\circ<\beta<23^\circ (95\% {\rm CL}) ~,
\nonumber\\
\gamma=(73\pm4)^\circ  & \mbox{ or } & 67^\circ<\gamma<81^\circ (95\% {\rm CL})
\end{eqnarray}
for Scheme (2B).

\FIGURE{
\centering
\includegraphics[width=7.4cm]{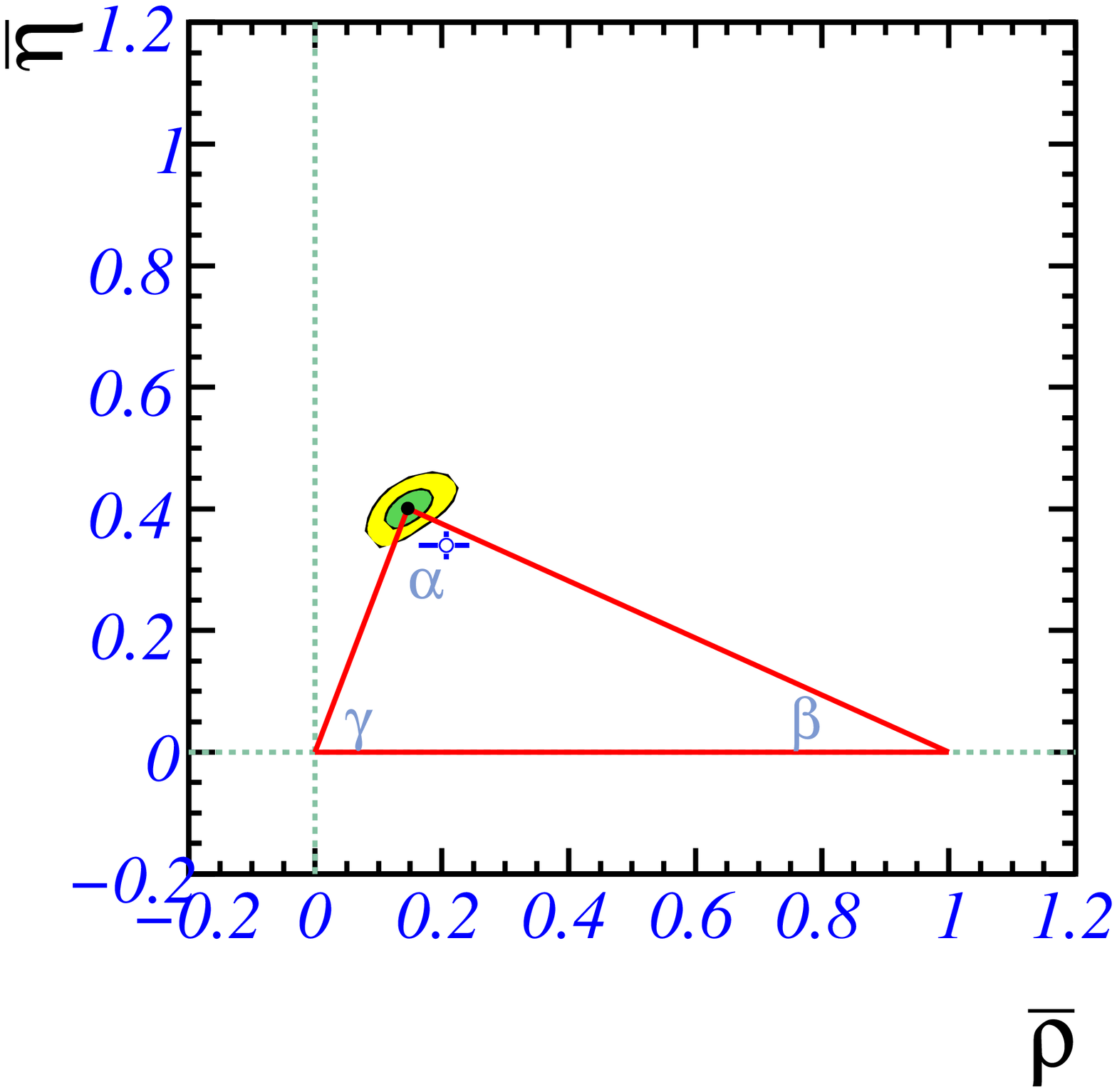}
\includegraphics[width=7.4cm]{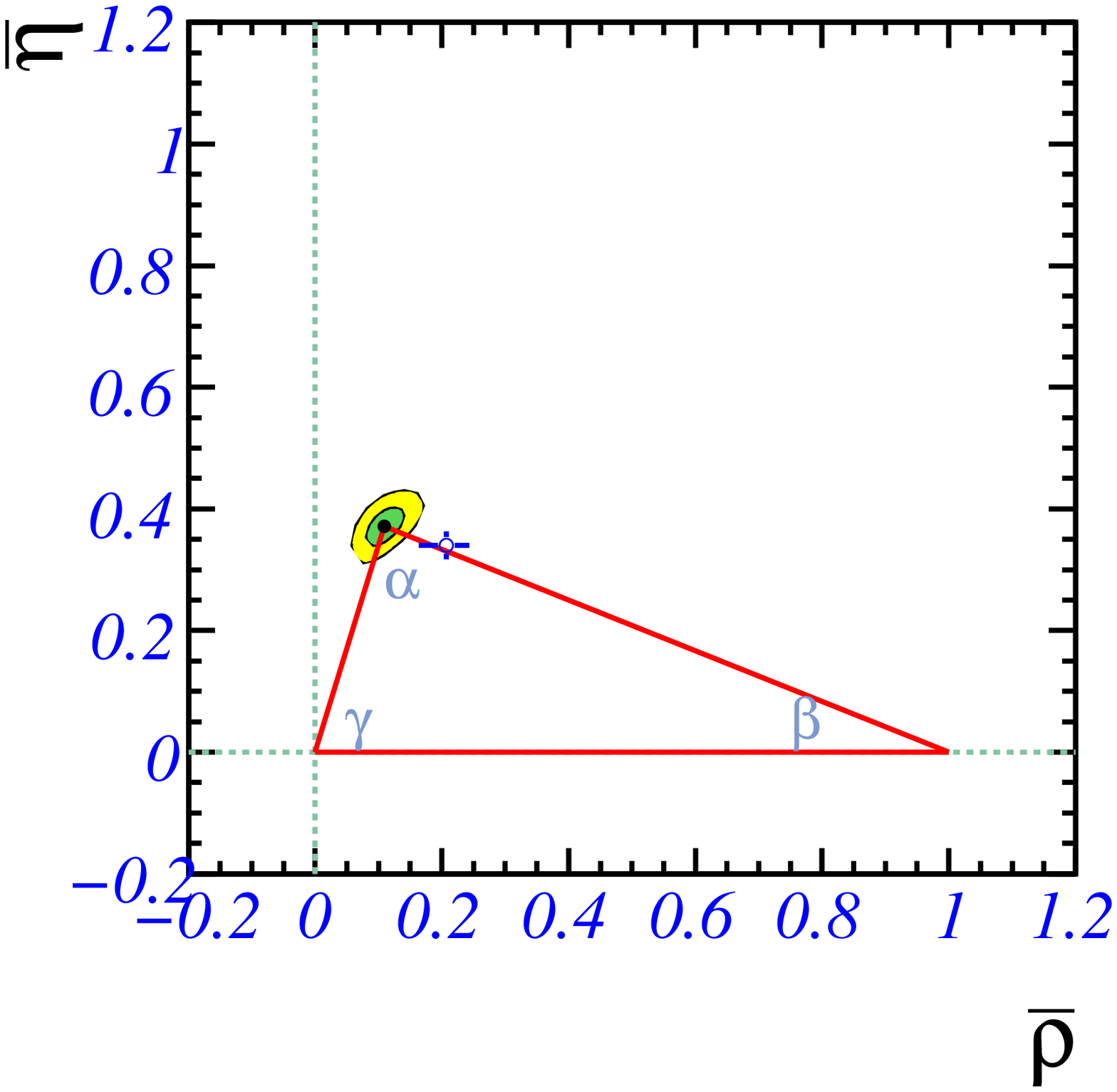}
\caption{The 1-$\sigma$ and 95\% CL contours of the $(\bar\rho, \bar\eta)$
  vertex obtained from a fit using the observed $VP$ modes that do not involve
  the singlet penguin amplitudes (left) and using all of the observed $VP$ modes
  (right), both assuming the exact flavor SU(3) symmetry.  The 1-$\sigma$ range
  given by the CKMfitter is indicated by the cross.}
\label{fig:rhoeta-scheme1}
}

The best-fitted UT vertex from the $VP$ modes is highly consistent with the one
from the $PP$ modes.  When the fits do not involve singlet penguin amplitudes,
both $VP$ and $PP$ data favor a slightly larger $\gamma \simeq 70^\circ$ and a
larger $\beta\simeq 26^\circ$.  After including the modes involving the singlet
penguin amplitudes, the best fitted $\gamma$ becomes even larger while $\beta$
reduces to the value consistent with the $B\to (c{\bar c}) K_S$ measurements.

\section{Discussions \label{sec:discuss}}

There are two sets of decay modes that can provide a good test for the SU(3)
symmetry.  One set contains the $B^+ \to K^{*0} \pi^+$, $B^+ \to {\lbar K}^{*0}
K^+$, $B^0 \to {\lbar K}^{*0} K^0$, and $B_s \to K^{*0} {\lbar K}^0$ modes.  The
other set contains the $B^+ \to \rho^+ K^0$, $B^+ \to K^{*+} {\lbar K}^0$, $B^0
\to K^{*0} {\lbar K}^0$, and $B_s \to {\lbar K}^{*0} K^0$ modes.  They all
involve only the $P_P$ or $P_V$ amplitude, where we have neglected the
$P_{EW,P}^C$ or $P_{EW,V}^C$ amplitude in the analysis as said before.  However,
this argument still applies if the color-suppressed EW penguin amplitude is
included because it scales in the same way as the QCD penguin amplitude.
Currently, only the $B^+ \to K^{*0} \pi^+$ and $B^+ \to \rho^+ K^0$ modes are
observed, and their branching ratios are measured at ${\cal O}(10^{-5})$ level.
It is thus very helpful to measure any of the $K^* K$ modes in this respect.
Using the fit results in Scheme (2A) and in units of $10^{-6}$, we predict the
branching ratios for the first set to be $10.64 \pm 0.82$, $0.50 \pm 0.05$,
$0.47 \pm 0.05$, and $9.11 \pm 0.70$, respectively.  The branching ratios for
the second set are $6.08 \pm 0.79$, $0.29 \pm 0.04$, $0.27 \pm 0.04$, and $5.21
\pm 0.68$ in units of $10^{-6}$, respectively.  These $B_{u,d} \to K^* K$ modes
are somewhat difficult to measure due to the Cabibbo suppression.  However, the
$B_s \to K^* K$ modes should be within the reach of the LHCb and Tevatron Run-II
experiments.

Although the $B^+ \to \phi \pi^+$ and $B^0 \to \phi \pi^0, \phi \eta, \phi
\eta'$ modes directly constrain the size of $s_P$, their branching ratios are
expected to be about ${\cal O}(10^{-8})$ or smaller.  Therefore, they are beyond
the current probes.

\TABLE{
  \caption{Predicted $B_{u,d,s}$ decay observables in Scheme (2A).  Numbers in
    the parentheses are the pulls of theory predictions from the current
    experimental data.}
\label{tab:Pred1A}
\begin{tabular}{|lllll|}
\hline
\multicolumn{2}{|c}{Mode} & BR ($\times 10^{-6}$) & $A_{CP}$ & ${\cal S}$ \\
\hline
$B_{u,d} \to$
& $\rho^-\pi^+         $  & $16.59\pm4.01 \ (-0.09)$              & $-0.042\pm0.041 \ (2.698)$            & $0.010\pm0.173 \ (-0.384)$\\              
& $\rho^+\pi^-         $  & $7.52 \pm1.97 \ (0.05)$               & $0.049\pm0.086 \ (-1.576)$            & $0.082\pm0.166 \ (-0.171)$\\              
& $\rho^0\pi^0         $  & $1.97 \pm0.94 \ (0.06)$               & $0.035\pm0.179 \ (-)$         & $-0.064\pm0.297 \ (-)$\\                          
& $\rho^+\pi^0         $  & $10.94\pm3.87 \ (-0.03)$              & $-0.011\pm0.193 \ (0.277)$            & $-$\\                                     
& $\rho^0\pi^+         $  & $8.81 \pm2.61 \ (-0.11)$              & $-0.121\pm0.090 \ (0.407)$            & $-$\\                                     
& $\bar{K}^{*0} K^0    $  & $0.47 \pm0.05 \ (-)$          & $0 \ (-)$         & $-$\\                                                     
& $K^{*0} \bar{K}^0    $  & $0.27 \pm0.04 \ (-)$          & $0 \ (-)$         & $-$\\                                                     
& $\bar{K}^{*0} K^+    $  & $0.50 \pm0.05 \ (0.94)$               & $0 \ (-)$         & $-$\\                                             
& $K^{*+} \bar{K}^0    $  & $0.29 \pm0.04 \ (-)$          & $0 \ (-)$         & $-$\\                                                     
& $\rho^- K^+          $  & $8.89 \pm1.13 \ (-0.29)$              & $0.094\pm0.094 \ (0.926)$             & $-$\\                                     
& $\rho^0 K^0          $  & $5.65 \pm1.21 \ (-0.26)$              & $0.076\pm0.031 \ (-0.331)$            & $0.824\pm0.047 \ (-0.822)$\\              
& $\rho^+ K^0          $  & $6.08 \pm0.79 \ (1.33)$               & $0 \ (-0.706)$            & $-$\\                                     
& $\rho^0 K^+          $  & $3.80 \pm0.96 \ (0.03)$               & $0.382\pm0.119 \ (0.401)$             & $-$\\                                     
& $K^{*0}\pi^0         $  & $6.59 \pm3.85 \ (-5.99)$              & $-0.330\pm0.120 \ (1.500)$            & $-$\\                                     
& $K^{*+}\pi^-         $  & $8.87 \pm0.76 \ (1.30)$               & $-0.043\pm0.075 \ (-1.882)$           & $-$\\                                     
& $K^{*0}\pi^+         $  & $10.64\pm0.82 \ (-0.80)$              & $0 \ (-0.339)$            & $-$\\                                     
& $K^{*+}\pi^0         $  & $7.00 \pm4.49 \ (-0.04)$              & $-0.081\pm0.272 \ (0.418)$            & $-$\\                                     
\hline
$B_s \to$
& $\rho^-K^+           $  & $6.89 \pm1.81 \ (-)$          & $0.049\pm0.086 \ (-)$         & $-$\\                                                     
& $\rho^0\bar{K}^0     $  & $0.39 \pm0.07 \ (-)$          & $0.929\pm0.195 \ (-)$         & $-0.357\pm0.528 \ (-)$\\                                  
& $K^{*-}\pi^+         $  & $15.22\pm3.68 \ (-)$          & $-0.042\pm0.041 \ (-)$                & $-$\\                                             
& $\bar{K}^{*0}\pi^0   $  & $2.60 \pm1.25 \ (-)$          & $-0.134\pm0.328 \ (-)$                & $-$\\                                             
& $K^{*-}K^+           $  & $7.45 \pm0.93 \ (-)$          & $0.085\pm0.084 \ (-)$         & $-$\\                                                     
& $K^{*+}K^-           $  & $8.16 \pm0.70 \ (-)$          & $-0.041\pm0.072 \ (-)$                & $-$\\                                             
& $\bar{K}^{*0} K^0    $  & $5.21 \pm0.68 \ (-)$          & $0 \ (-)$         & $-$\\                                                     
& $K^{*0}\bar{K}^0     $  & $9.11 \pm0.70 \ (-)$          & $0 \ (-)$         & $-$\\                                               
\hline
\end{tabular}
}

In the following, we would like to point out some persistent problems
encountered in our fits to the current data.  In Ref.~\cite{Chiang:2003pm}, the
rate difference relations \cite{Deshpande:2000jp}:
\begin{eqnarray}
\label{eq:check1}
\Gamma(B^0 \to \rho^- \pi^+) - \Gamma({\bar B}^0 \to \rho^+ \pi^-)
&=&
\frac{f_\pi}{f_K}
\left[ \Gamma({\bar B}^0 \to \rho^+ K^-) - \Gamma(B^0 \to \rho^- K^+) \right] ~,
 \\
\label{eq:check2}
\Gamma(B^0 \to \rho^+ \pi^-) - \Gamma({\bar B}^0 \to \rho^- \pi^+)
&=&
\frac{f_\rho}{f_{K^*}}
\left[ \Gamma({\bar B}^0 \to K^{*-} \pi^+) - \Gamma(B^0 \to K^{*+} \pi^-)
\right]
\nonumber \\
\end{eqnarray}
have been found to be barely and loosely obeyed, respectively, by the data at
that time.  Using the current data and in terms of the branching ratios,
Eqs.~(\ref{eq:check1}) and (\ref{eq:check2}) give in units of $10^{-6}$,
respectively,
\begin{eqnarray}
-3.9 \pm 2.0
&\stackrel{?}{=}&
2.1 \pm 0.9 ~, \\
2.1 \pm 1.9
&\stackrel{?}{=}&
-4.9 \pm 2.2 ~.
\end{eqnarray}
The first one is still not obeyed at about $2.7\sigma$ level.  This difference
comes from the CP asymmetries of $B^0 \to \rho^- \pi^+$ and $\rho^- K^+$, both
at about $2 \sigma$ level.  To further check the equality in the second equation
relies on more precise determinations in the CP asymmetries of $B^0 \to K^{*+}
\pi^-$ and $B^0 \to K^{*+} \pi^-$.

Another problem is ${\cal B}(B^+ \to \rho^+ \eta') / {\cal B}(B^+ \to \rho^+
\eta) \simeq 1.3 \pm 0.5$, which is very different from our expectation of about
$1/2$ based upon the mixing angle we assume for $\eta$ and $\eta'$ and assuming
that $s_V$ is negligible for $\Delta S = 0$ decays.  The problem comes from the
large branching ratio of $B^+ \to \rho^+ \eta'$, as indicated by the pull in
Table~\ref{tab:PredictS1}.  A similar relation can be found for ${\cal B}(B^0
\to \rho^0 \eta) / {\cal B}(B^0 \to \rho^0 \eta')$, ${\cal B}(B^0 \to \omega^0
\eta) / {\cal B}(B^0 \to \omega^0 \eta')$, ${\cal B}(B_s \to \rho^0 \eta) /
{\cal B}(B_s \to \rho^0 \eta')$, and ${\cal B}(B_s \to \omega \eta) / {\cal
  B}(B_s \to \omega \eta')$ too.  However, these modes may be difficult to
measure.

A new problem would occur between the $B^+ \to \rho^0 K^+$ and $\omega K^+$
modes that differ by $\sqrt{2} s'_P$ if $s'_P$ is vanishingly small.  In that
case, the ratio of their branching ratios should be close to 1
\cite{Lipkin:1998ew}.  However, the current data comes down to $0.57 \pm 0.08$.
With the fitted $S_P \simeq 140$ eV, the predicted ratio is $\simeq 0.61$.
Consequently, a non-vanishing $S_P$ is preferred.

Another puzzle comes from the CP asymmetry of $B^0 \to K^{*0} \eta$ because
it is measured at an almost $4\sigma$ level.  This is quite different from a
closely related mode, $B^+ \to K^{*+} \eta$, whose CP asymmetry is consistent
with zero.  Their values should not be so different because they only differ by
a small tree amplitude.

We make predictions for the observables of all the $B^+$, $B^0$ and $B_s$ decays
using the extracted parameters given in Table~\ref{tab:Fit}.  In
Table~\ref{tab:Pred1A}, we only include modes without involving the singlet
penguin amplitudes as they are based on Scheme (2A).  Table~\ref{tab:PredictS1}
and Table~\ref{tab:PredictS2} cover all the decay modes as they are based on
Scheme (2B).  The column of $A_{CP}$ refers to either the direct CP asymmetry or
${\cal A}$ in Eq.~(\ref{eq:t-dep}) of the corresponding mode.  The numbers in
the parentheses are calculated pulls of the theory predictions from experimental
observations.  They indicate the $\Delta \chi^2$ contributions of individual
quantities.  

Several observables in Table~\ref{tab:PredictS1} have pulls larger than, say
1.5.  Most of them are in the CP asymmetries.  It is less clear about their
importance as current precision on these data points is not satisfactory.  We
are then left with two branching ratio predictions with large pulls.  The
problem with $\rho^+ \eta'$ has been mentioned above.  As commented before, we
do not include the branching ratio and CP asymmetry of the $B^0 \to K^{*0}
\pi^0$ in the fits of this work.  Its predicted branching ratios in
Tables~\ref{tab:Pred1A} and \ref{tab:PredictS1} based on the best fits are quite
different from the current quotes of averages in Table~\ref{tab:VPdS1}, and need
further experimental confirmation.

In Table~\ref{tab:PredictS1}, our predictions of ${\cal B}(B^0 \to \rho^0 \eta)
= 1.87 \pm 0.64$ and ${\cal B}(B^0 \to \omega \pi^0) = 2.82 \pm 0.99$ are larger
than the current upper bounds of $1.5$ and $0.5$, respectively, in units of
$10^{-6}$.  The branching ratio predictions of the other yet-measured modes are
all consistent with current $95\%$ upper bounds.

For the $B_s$ decays, we predict large direct CP asymmetries
$A_{CP}(\overline{K}^{*0} \eta)\simeq 0.73$ and $A_{CP}(\overline{K}^{*0}
\eta')\simeq -0.79$, a result of interference between the large color-suppressed
amplitude $C_V$ and the QCD penguin amplitudes.  We also predict large branching
ratios, in unit of $10^{-6}$, ${\cal B}(\phi \eta')\simeq 8.47$, ${\cal
  B}(K^{*-} \pi^+)\simeq 15.21$, ${\cal B}(K^{*\pm}K^\mp)\simeq 8$, and ${\cal
  B}(K^{*0} \overline{K}^0)\simeq 9.54$. In these modes, the branching ratios
can reach ${\cal O}(10^{-5})$ or more, as they involve either $T_V$ for $\Delta
S = 0$ or $P_P$ for $|\Delta S| = 1$ transitions.

\TABLE{
  \caption{Predicted $B_{u,d}$ decay observables in Scheme (2B).  Numbers in the
    parentheses are the pulls of theory predictions from the current
    experimental data.}
\label{tab:PredictS1}
\begin{tabular}{|llll|}
\hline
Mode & BR ($\times 10^{-6}$) & $A_{CP}$ & ${\cal S}$ \\
\hline
$\rho^-\pi^+         $  & $16.57\pm4.18 \ (-0.08)$              & $-0.038\pm0.041 \ (2.630)$            & $0.070\pm0.166 \ (-0.843)$\\                 
$\rho^+\pi^-         $  & $7.32 \pm1.98 \ (0.21)$               & $0.024\pm0.072 \ (-1.363)$            & $0.084\pm0.160 \ (-0.187)$\\                 
$\rho^0\pi^0         $  & $1.91 \pm0.79 \ (0.19)$               & $0.259\pm0.148 \ (-)$         & $0.115\pm0.249 \ (-)$\\                              
$\rho^+\pi^0         $  & $11.12\pm2.99 \ (-0.15)$              & $-0.026\pm0.128 \ (0.415)$            & $-$\\                                        
$\rho^0\pi^+         $  & $8.27 \pm2.42 \ (0.41)$               & $-0.192\pm0.099 \ (0.977)$            & $-$\\                                        
$\rho^0\eta          $  & $1.87 \pm0.64 \ (-)$          & $0.109\pm0.153 \ (-)$         & $-0.336\pm0.199 \ (-)$\\                                     
$\rho^0\eta'         $  & $0.52 \pm0.15 \ (-)$          & $-0.396\pm0.291 \ (-)$                & $-0.587\pm0.222 \ (-)$\\                             
$\rho^+\eta          $  & $7.16 \pm2.03 \ (-0.26)$              & $0.165\pm0.103 \ (-0.502)$            & $-$\\                                        
$\rho^+\eta'         $  & $3.79 \pm0.98 \ (1.63)$               & $-0.071\pm0.240 \ (0.110)$            & $-$\\                                        
$\omega\pi^0         $  & $2.82 \pm0.99 \ (-)$          & $0.293\pm0.132 \ (-)$         & $-0.094\pm0.216 \ (-)$\\                                     
$\omega\pi^+         $  & $7.02 \pm2.23 \ (-0.25)$              & $0.020\pm0.075 \ (-0.993)$            & $-$\\                                        
$\omega\eta          $  & $1.27 \pm0.51 \ (-)$          & $-0.016\pm0.179 \ (-)$                & $-0.360\pm0.227 \ (-)$\\                             
$\omega\eta'         $  & $0.76 \pm0.25 \ (-)$          & $-0.624\pm0.285 \ (-)$                & $-0.511\pm0.302 \ (-)$\\                             
$\phi\pi^0           $  & $0.02 \pm0.01 \ (-)$          & $0 \ (-)$         & $0 \ (-)$\\                              
$\phi\pi^+           $  & $0.04 \pm0.02 \ (-)$          & $0 \ (-)$         & $-$\\                                                
$\phi\eta            $  & $0.01 \pm0.01 \ (-)$          & $0 \ (-)$         & $0 \ (-)$\\                              
$\phi\eta'           $  & $0.01 \pm0.00 \ (-)$          & $0 \ (-)$         & $0 \ (-)$\\                              
$\bar{K}^{*0} K^0    $  & $0.52 \pm0.05 \ (-)$          & $0 \ (-)$         & $-$\\                                                
$K^{*0} \bar{K}^0    $  & $0.31 \pm0.04 \ (-)$          & $0 \ (-)$         & $-$\\                                                
$\bar{K}^{*0} K^+    $  & $0.55 \pm0.05 \ (0.67)$               & $0 \ (-)$         & $-$\\                                        
$K^{*+} \bar{K}^0    $  & $0.33 \pm0.04 \ (-)$          & $0 \ (-)$         & $-$\\                                                
$\rho^- K^+          $  & $9.21 \pm1.04 \ (-0.61)$              & $0.082\pm0.089 \ (1.128)$             & $-$\\                                        
$\rho^0 K^0          $  & $5.06 \pm1.10 \ (0.36)$               & $-0.041\pm0.045 \ (0.072)$            & $0.766\pm0.052 \ (-0.598)$\\                 
$\rho^+ K^0          $  & $6.70 \pm0.74 \ (0.90)$               & $0 \ (-0.706)$            & $-$\\                                
$\rho^0 K^+          $  & $4.02 \pm0.82 \ (-0.44)$              & $0.382\pm0.126 \ (0.398)$             & $-$\\                                        
$\omega \bar{K}^0    $  & $4.62 \pm1.01 \ (0.63)$               & $0.033\pm0.048 \ (1.690)$             & $0.700\pm0.054 \ (-1.040)$\\                 
$\omega K^+          $  & $6.64 \pm1.27 \ (0.13)$               & $0.029\pm0.092 \ (-0.190)$            & $-$\\                                        
$\phi K^0            $  & $7.43 \pm1.21 \ (0.79)$               & $0 \ (1.533)$             & $0.737\pm0.043 \ (-1.699)$\\         
$\phi K^+            $  & $7.96 \pm1.30 \ (0.53)$               & $0 \ (0.773)$             & $-$\\                                
$K^{*0}\pi^0         $  & $13.85\pm4.76 \ (-16.36)$             & $-0.294\pm0.078 \ (1.201)$            & $-$\\                                        
$K^{*+}\pi^-         $  & $9.57 \pm0.72 \ (0.66)$               & $-0.019\pm0.057 \ (-2.104)$           & $-$\\                                        
$K^{*0}\pi^+         $  & $11.14\pm0.77 \ (-1.43)$              & $0 \ (-0.339)$            & $-$\\                                
$K^{*+}\pi^0         $  & $7.09 \pm3.11 \ (-0.08)$              & $-0.151\pm0.164 \ (0.660)$            & $-$\\                                        
$K^{*0}\eta          $  & $16.72\pm2.44 \ (-0.82)$              & $0.162\pm0.049 \ (0.560)$             & $-$\\                                        
$K^{*0}\eta'         $  & $4.16 \pm1.56 \ (-0.30)$              & $0.159\pm0.150 \ (-0.954)$            & $-$\\                                        
$K^{*+}\eta          $  & $17.30\pm2.58 \ (1.25)$               & $0.070\pm0.064 \ (-0.837)$            & $-$\\                                        
$K^{*+}\eta'         $  & $4.34 \pm1.64 \ (0.28)$               & $-0.027\pm0.228 \ (0.933)$            & $-$\\                                        
\hline
\end{tabular}
}

\TABLE{
  \caption{Predicted $B_s$ decay observables in Scheme (2B).  Numbers in the
    parentheses are the pulls of theory predictions from the current
    experimental data.}
\label{tab:PredictS2}
\begin{tabular}{|llll|}
\hline
Mode & BR ($\times 10^{-6}$) & $A_{CP}$ & ${\cal S}$ \\
\hline
$\rho^0\eta          $  & $0.21 \pm0.14 \ (-)$          & $-0.156\pm0.123 \ (-)$                & $-0.731\pm0.092 \ (-)$\\     
$\rho^0\eta'         $  & $0.42 \pm0.26 \ (-)$          & $-0.156\pm0.123 \ (-)$                & $-0.731\pm0.092 \ (-)$\\     
$\rho^-K^+           $  & $6.71 \pm1.81 \ (-)$          & $0.024\pm0.072 \ (-)$         & $-$\\                                
$\rho^0\bar{K}^0     $  & $0.24 \pm0.10 \ (-)$          & $-0.128\pm0.773 \ (-)$                & $0.926\pm0.283 \ (-)$\\      
$\omega \eta         $  & $0.07 \pm0.06 \ (-)$          & $0.243\pm0.234 \ (-)$         & $-0.624\pm0.195 \ (-)$\\             
$\omega \eta'        $  & $0.13 \pm0.12 \ (-)$          & $0.243\pm0.234 \ (-)$         & $-0.624\pm0.195 \ (-)$\\             
$\omega \bar{K}^0    $  & $0.27 \pm0.14 \ (-)$          & $0.302\pm0.629 \ (-)$         & $-0.856\pm0.331 \ (-)$\\             
$\phi\pi^0           $  & $2.80 \pm1.80 \ (-)$          & $-0.250\pm0.121 \ (-)$                & $-0.451\pm0.131 \ (-)$\\     
$\phi\eta            $  & $2.35 \pm1.53 \ (-)$          & $-0.073\pm0.142 \ (-)$                & $-0.341\pm0.174 \ (-)$\\     
$\phi\eta'           $  & $8.47 \pm2.55 \ (-)$          & $0.096\pm0.061 \ (-)$         & $-0.626\pm0.054 \ (-)$\\             
$\phi\bar{K}^0       $  & $0.44 \pm0.07 \ (-)$          & $0 \ (-)$         & $0 \ (-)$\\              
$K^{*-}\pi^+         $  & $15.21\pm3.83 \ (-)$          & $-0.038\pm0.041 \ (-)$                & $-$\\                        
$\bar{K}^{*0}\pi^0   $  & $4.27 \pm1.36 \ (-)$          & $-0.064\pm0.146 \ (-)$                & $-$\\                        
$\bar{K}^{*0}\eta    $  & $3.26 \pm0.93 \ (-)$          & $0.730\pm0.108 \ (-)$         & $-$\\                                
$\bar{K}^{*0}\eta'   $  & $1.99 \pm0.47 \ (-)$          & $-0.794\pm0.191 \ (-)$                & $-$\\                        
$K^{*-}K^+           $  & $7.79 \pm0.86 \ (-)$          & $0.073\pm0.079 \ (-)$         & $-$\\                                
$K^{*+}K^-           $  & $8.79 \pm0.66 \ (-)$          & $-0.018\pm0.054 \ (-)$                & $-$\\                        
$\bar{K}^{*0} K^0    $  & $5.74 \pm0.63 \ (-)$          & $0 \ (-)$         & $-$\\                                
$K^{*0}\bar{K}^0     $  & $9.54 \pm0.66 \ (-)$          & $0 \ (-)$         &
$-$\\  
\hline
\end{tabular}
}

\section{Summary \label{sec:summary}}

We have updated the global analysis of charmless $B \to VP$ decays in the
framework of flavor SU(3) symmetry using the latest experimental data.
Moreover, we consider different SU(3) breaking schemes for the sizes of flavor
amplitudes based upon factorization assumption.  Our result shows that the
symmetry-breaking scheme (Scheme 2 defined in the text) is favored by the
$\chi^2$ fits, but its difference from the exact symmetry scheme (Scheme 1) is
small.  The UT vertex $(\bar\rho,\bar\eta)$ extracted using these modes is
consistent with our previous analysis using the $PP$ modes \cite{Chiang:2006ih},
and also agrees with other methods within errors \cite{CKMfitter,UTfit}.
However, we note that a slightly larger weak phase $\gamma$ is favored by our
global analysis.

In the fits to modes without involving the singlet penguin amplitudes, we note
that there are two sets of solutions with minimal $\chi^2$ values.  In one set,
the $P_P$ and $P_V$ amplitudes have almost the same strong phases.  In the other
set, they have almost opposite strong phases.  The latter is favored when one
also includes modes involving the singlet penguin amplitudes.  Moreover, we find
in the latter case that the ratio $C_V/T_V$ is about $0.6$ - $0.7$, similar to
the $C/T$ ratio in the $PP$ modes.  Correspondingly, the $P_{EW,V}$ and $S_V$
amplitudes are unexpectedly large.  These facts are seen to be a challenge to
perturbative approaches.

We point out that a set of decay modes that involve only the QCD penguin
amplitude can be used to test our flavor SU(3) assumption.  Among those modes,
the $B_s \to K^{*0} \overline{K}^0$ and $\overline{K}^{*0} K^0$ modes should be
within the reach of the LHCb and Tevatron Run-II experiments.

We also mention the persistent problems that the CP rate differences in $B^0 \to
\rho^- \pi^+$ and in $B^0 \to \rho^- K^+$ do not follow our expectation from
factorization and that the observed branching ratio of $B^+ \to \rho^+ \eta'$ is
too large to be accommodated in our approach.  Further investigations of ${\cal
  B} (B^0 \to K^{*0} \pi^0)$ and $A_{CP}(B^0 \to K^{*0} \eta)$ are required.

Based on our best fits, we calculate all observables in the $B \to VP$ decays.
The part for $B_s$ decays is particularly useful because currently no such
observables have been observed yet and our results serve as predictions to be
compared with.

\bigskip

\section*{Acknowledgments}

The authors would like to thank the hospitality of Kavli Institute of
Theoretical Physics in Beijing where part of this work is done.  We also
appreciate useful discussions and comments from I.~Bigi, X.-G.~He, H.-n.~Li, and
C.~Sachrajda and the information on the latest ICHEP data from P.~Chang and
J.~Smith.  C.~C. would like to thank the hospitality of the National Center for
Theoretical Sciences in Hsinchu, where part of this work is done.  This research
was supported in part by the National Science Council of Taiwan, R.~O.~C.\ under
Grant No.\ NSC 96-2112-M-008-001.


\end{document}